\RequirePackage{fix-cm}
\documentclass[twocolumn]{svjour3}          % twocolumn
\smartqed  % flush right qed marks, e.g. at end of proof
\usepackage{lipsum,mathtools,cuted}
\usepackage{graphicx}
\usepackage{bbding}
\usepackage{upgreek}
\usepackage{url}
\usepackage[dvipsnames]{color,xcolor}
\usepackage{amsmath}
\usepackage{bm}
\usepackage[colorlinks,
            linkcolor=blue,
            anchorcolor=blue,
            citecolor=blue,
            urlcolor=blue,
            ]{hyperref}
\usepackage{mathptmx}      % use Times fonts if available on your TeX system
\usepackage{titletoc}
\usepackage{amssymb}
\usepackage{flushend}
\usepackage{multirow}
\usepackage{caption}
\usepackage{subfig}

\newcommand\norm[1]{\left\lVert#1\right\rVert}

\emergencystretch=\maxdimen
\titlecontents{section}[0pt]{\addvspace{2pt}\filright}              {\contentspush{\thecontentslabel\ }}              {}{\titlerule*[8pt]{.}\contentspage}

\hypersetup{
    bookmarks=false,         % show bookmarks bar?
    colorlinks=true,       % false: boxed links; true: colored links
    linkcolor=blue,          % color of internal links (change box color with linkbordercolor)
    linkbordercolor={1 1 1}
}

\journalname{Acta Mech. Sin.}
\begin{document}
%\begin{sloppypar}
\rmfamily
\title{Physics-informed neural networks (PINNs) for fluid mechanics: A review %\thanks{Grants or other notes
}
%\subtitle{({\it Acta Mechanica Sinica})}

\titlerunning{Physics-informed neural networks (PINNs) for fluid mechanics: A review}        % if too long for running head

\author{          Shengze Cai$^1$ \and Zhiping Mao$^2$ \and Zhicheng Wang$^3$ \and  Minglang Yin$^{4,5}$ \\
 George Em Karniadakis$^{1,4*}$%\and
       }


\institute{{\Envelope} George Em Karniadakis \at
            \email{george\_karniadakis@brown.edu} \at \at
%            {\Envelope} corAuthor \at
%             \email{actams@cstam.org.cn} \at \at
		    $^{1}$ Division of Applied Mathematics, Brown University, Providence, RI 02912, USA \at
            $^{2}$ School of Mathematical Sciences Xiamen University, 361005 China  \at
            $^{3}$ Laboratory of Ocean Energy Utilization of Ministry of Education, Dalian University of Technology, Dalian, 116024, China  \at
            $^{4}$ School of Engineering, Brown University, Providence, RI 02912, USA  \at
            $^{5}$ Center for Biomedical Engineering, Brown University, Providence, RI 02912, USA
           }
%Received: date / Accepted: date / Published online: 7 February 2020\\
\date{\copyright {\it Acta Mechanica Sinica}, The Chinese Society of Theoretical and Applied Mechanics (CSTAM) 2020
}

%\contentsline{section}{\numberline{12.8}Some section that is wrapped in the TOC}{87}

\maketitle

%\tableofcontents

\begin{abstract}
Despite the significant progress over the last 50 years in simulating flow problems using numerical discretization of the Navier-Stokes equations (NSE),  we still cannot incorporate seamlessly noisy data into existing algorithms, mesh-generation is complex, and we cannot tackle high-dimensional problems governed by parametrized NSE. Moreover, solving {\em inverse flow problems} is often prohibitively expensive and requires complex and expensive formulations and new computer codes. Here, we review {\em flow physics-informed learning}, integrating seamlessly data and mathematical models, and implementing them using physics-informed neural networks (PINNs). We demonstrate the effectiveness of PINNs for inverse problems related to three-dimensional wake flows, supersonic flows, and biomedical flows. 
\keywords{Physics-informed learning \and PINNs \and Inverse problems \and Supersonic flows \and Biomedical flows}
\end{abstract}
\vspace{1 cm}

% ================================
\section{Introduction}

In the last 50 years there has been a tremendous progress in computational fluid dynamics (CFD) in solving numerically 
the incompressible and compressible Navier-Stokes equations (NSE) using finite elements, spectral, and even meshless methods \cite{brooks1982streamline,GK_CFDbook,katz2009meshless,liu2010smoothed}. Yet, for real-world applications, we still cannot incorporate seamlessly (multi-fidelity) data into existing algorithms, and for industrial-complexity problems the mesh generation is time consuming and still an art. Moreover, solving {\em inverse problems}, e.g., for unknown boundary conditions or conductivities~\cite{beck1985inverse}, etc., is often prohibitively expensive and requires different formulations and new computer codes. Finally, computer programs such as OpenFOAM~\cite{jasak2007openfoam} have more than 100,000 lines of code, making it almost impossible to maintain and update them from one generation to the next. 

Physics-informed learning~\cite{raissi2021physics}, introduced in a series of papers by Karniadakis's group both for Gaussian-process regression~\cite{raissi2017machine,raissi2018numerical} and physics-informed neural networks (PINNs)~\cite{raissi2019deep}, can seamlessly integrate multifidelity/multimodality experimental data with the various Navier-Stokes formulations for incompressible flows~\cite{jin2021nsfnets,raissi2020hidden} as well as compressible flows~\cite{mao2020physics} and biomedical flows~\cite{yin2021non}. PINNs use automatic differentiation to represent all the differential operators and hence there is no explicit need for a mesh generation. Instead, the Navier-Stokes equations and any other kinematic or thermodynamic constraints can be directly incorporated in the loss function of the neural network (NN) by penalizing deviations from the target values (e.g., zero residuals for the conservation laws) and are properly weighted with any given data, e.g., partial measurements of the surface pressure. PINNs are not meant to be a replacement of the existing CFD codes, and in fact the current generation of PINNs is not as accurate or as efficient as high-order CFD codes~\cite{GK_CFDbook} for solving the standard forward problems. This limitation is associated with the minimization of the loss function, which is a high-dimensional non-convex function, a limitation which is a grand challenge of all neural networks for even commercial machine learning. However, PINNs perform much more accurately and more efficiently than any CFD solver if any scattered partial spatio-temporal data are available for the flow problem under consideration. Moreover, the forward and inverse PINN formulations are identical so there is no need for expensive data assimilation schemes that have stalled progress especially for optimization and design applications of flow problems in the past.  

In this paper we first review the basic principles of PINNs and recent extensions using domain decomposition for multiphysics and multiscale flow problems. We then present new results for a three-dimensional (3D) wake formed in incompressible flow behind a circular cylinder. We also show results for a two-dimensional (2D) supersonic flow past a blunt body, and finally we infer material parameters in simulating thrombus deformation in a biomedical flow. 

% ================================
\section{PINNs: Physics-Informed Neural Networks}\label{sec:PINNs}

In this section we first review the basic PINN concept and subsequently discuss more recent advancements in incompressible, compressible and biomedical flows. 

\begin{figure*}[t]
    \centering
	\includegraphics[width=0.8\textwidth]{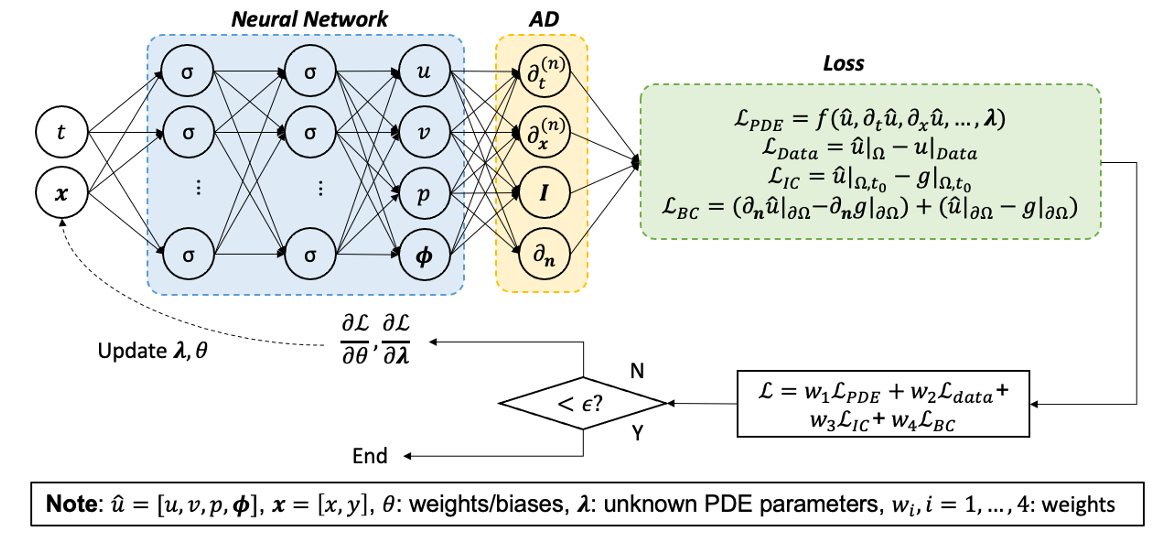} 
    \caption{\textbf{Schematic of a physics-informed neural network (PINN).} A fully-connected neural network, with time and space coordinates ($t,\mathbf{x}$) as inputs, is used to approximate the multi-physics solutions $\hat{u}=[u,v,p,\phi]$. The derivatives of $\hat{u}$ with respect to the inputs are calculated using automatic differentiation (AD) and then used to formulate the residuals of the governing equations in the loss function, that is generally composed of multiple terms weighted by different coefficients. The parameters of the neural network $\theta$ and the unknown PDE parameters $\lambda$ can be learned simultaneously by minimizing the loss function. 
    }
    \label{fig:schematic}
\end{figure*}

\subsection{PINNs: Basic Concepts}

We consider a parametrized partial differential equation (PDE) system given by:
\begin{equation}
\begin{aligned}
&f(\mathbf{x},t,\hat{u},\partial_{\mathbf{x}}\hat{u},\partial_{t}\hat{u},\dots, \bm{\lambda} ) = 0, \quad \mathbf{x}\in \Omega, ~ t\in [0,T] \\
&\hat{u}(\mathbf{x},t_{0}) = g_{0}(\mathbf{x}) \quad \mathbf{x}\in \Omega, \\
&\hat{u}(\mathbf{x},t) = g_{\Gamma}(t), \quad \mathbf{x}\in \partial\Omega, ~ t\in [0,T],
\end{aligned}
\label{eq:PDE}
\end{equation}
where $\mathbf{x}\in \mathbb{R}^{d}$ is the spatial coordinate and $t$ is the time; $f$ denotes the residual of the PDE, containing the differential operators (i.e., $[\partial_{\mathbf{x}}\hat{u},\partial_{t}\hat{u},\dots]$); 
$\bm{\lambda}=[\lambda_{1},\lambda_{2},\dots]$ are the PDE parameters; $\hat{u}(\mathbf{x},t)$ is the solution of the PDE with initial condition $g_{0}(\mathbf{x})$ and boundary condition $g_{\Gamma}(t)$ (which can be Dirichlet, Neumann or mixed boundary condition); $\Omega$ and $\partial \Omega$ represent the spatial domain and the boundary, respectively.  

In the context of the vanilla PINNs~\cite{raissi2019physics}, a fully-connected feed-forward neural network, which is composed of multiple hidden layers, is used to approximate the solution of the PDE $\hat{u}$ by taking the space and time coordinates $(\mathbf{x},t)$ as inputs, as shown in the blue panel in Fig.~\ref{fig:schematic}. Let the hidden variable of the $k^{th}$ hidden layer be denoted by $\mathbf{z}^{k}$, then the neural network can be expressed as 
\begin{equation}
\begin{aligned}
&\mathbf{z}^{0} = (\mathbf{x},t), \\
&\mathbf{z}^{k} = \sigma(\mathbf{W}^{k}\mathbf{z}^{k-1} + \mathbf{b}^{k} ), \quad 1 \le k \le L-1 \\
&\mathbf{z}^{k} = \mathbf{W}^{k}\mathbf{z}^{k-1} + \mathbf{b}^{k} , \quad k= L,
\end{aligned}
\label{eq:FNN}
\end{equation}
where the output of the last layer is used to approximate the true solution, namely $\hat{u} \approx \mathbf{z}^{L}$. $\mathbf{W}^{k}$ and $\mathbf{b}^{k}$ denote the weight matrix and bias vector of the $k^{th}$ layer; $\sigma(\cdot)$ is a nonlinear activation function. All the trainable model parameters, i.e., weights and biases, are denoted by $\theta$ in this paper. 

In PINNs, solving a PDE system (denoted by equ. \ref{eq:PDE}) is converted into an optimization problem by iteratively updating $\theta$ with the goal to minimize the loss function $L$:
\begin{equation}
L = \omega_{1}L_{PDE} + \omega_{2}L_{data} + \omega_{3}L_{IC} + \omega_{4}L_{BC},
\label{eq:PINN_loss}
\end{equation}
where $\omega_{1-4}$ are the weighting coefficients for different loss terms. The first term $L_{PDE}$ in equ. \ref{eq:PINN_loss} penalizes the residual of the governing equations. The other terms are imposed to satisfy the model predictions for the measurements $L_{data}$, the initial condition $L_{IC}$, and the boundary condition $L_{BC}$, respectively. In general, the mean square error (MSE), taking the $L_2$-norm of the sampling points, is employed to compute the losses in equ. \ref{eq:PINN_loss}. The sampling points are defined as a data set $\{\mathbf{x}^{i},t^{i}\}_{i=1}^{N}$, where the number of points (denoted by $N$) for different loss terms can be different. Generally, we use the ADAM optimizer~\cite{kingma2014adam}, an adaptive algorithm for gradient-based first-order optimization, to optimize the model parameters $\theta$.  

\textbf{Remark 1:} We note that the definition of the loss function shown in equ. \ref{eq:PINN_loss} is problem-dependent, hence some terms may disappear for different types of the problem. For example, when we solve a forward problem in fluid mechanics with the known parameters ($\mathbf{\lambda}$) and the initial/boundary conditions of the PDEs, the data loss $L_{data}$ is not necessarily required. However, in the cases where the model parameters or the initial/boundary conditions are unknown (namely, inverse problems), the data measurements should be taken into account in order to make the optimization problem solvable.
We also note that the PINN framework can be employed to solve an ``over-determined" system, e.g., well-posed in a classical sense with initial and boundary conditions known and additionally some measurements inside the domain or at boundaries (e.g., pressure measurements).

One of the key procedures to construct the PDE loss in equ. \ref{eq:PINN_loss} is the computation of partial derivatives, which is addressed by using automatic differentiation (AD). Relying on the combination of the derivatives for a sequence of  operations by using the chain rule, AD calculates the derivatives of the outputs with respect to the network inputs directly in the computational graph. 
The computation of partial derivatives can be calculated with an explicit expression, hence avoiding introducing truncation errors in conventional numerical approximations. At the present time, AD has been implemented in various deep learning frameworks~\cite{abadi2016tensorflow,paszke2019pytorch}, which makes it convenient for the development of PINNs. 

A schematic of PINNs is shown in Fig.~\ref{fig:schematic}, where the key elements (e.g., neural network, AD, loss function) are indicated in different colors. Here, we consider a multi-physics problem, where the solutions include the velocity $(u,v)$, pressure $p$ and a scalar field $\phi$, which are coupled in a PDE system $f$. The schematic in Fig.~\ref{fig:schematic} represents most of the typical problems in fluid mechanics. For instance, the PDEs considered here can be the Boussinesq approximation of the Navier-Stokes equations, where $\phi$ is the temperature. Following the paradigm in Fig.~\ref{fig:schematic}, we will describe the governing equations, the loss function and the neural network configurations of PINNs case-by-case in the rest of this paper. 
% ==================================
% Figure 2
% ==================================
\begin{figure*}[t]
  \includegraphics[width=\textwidth]{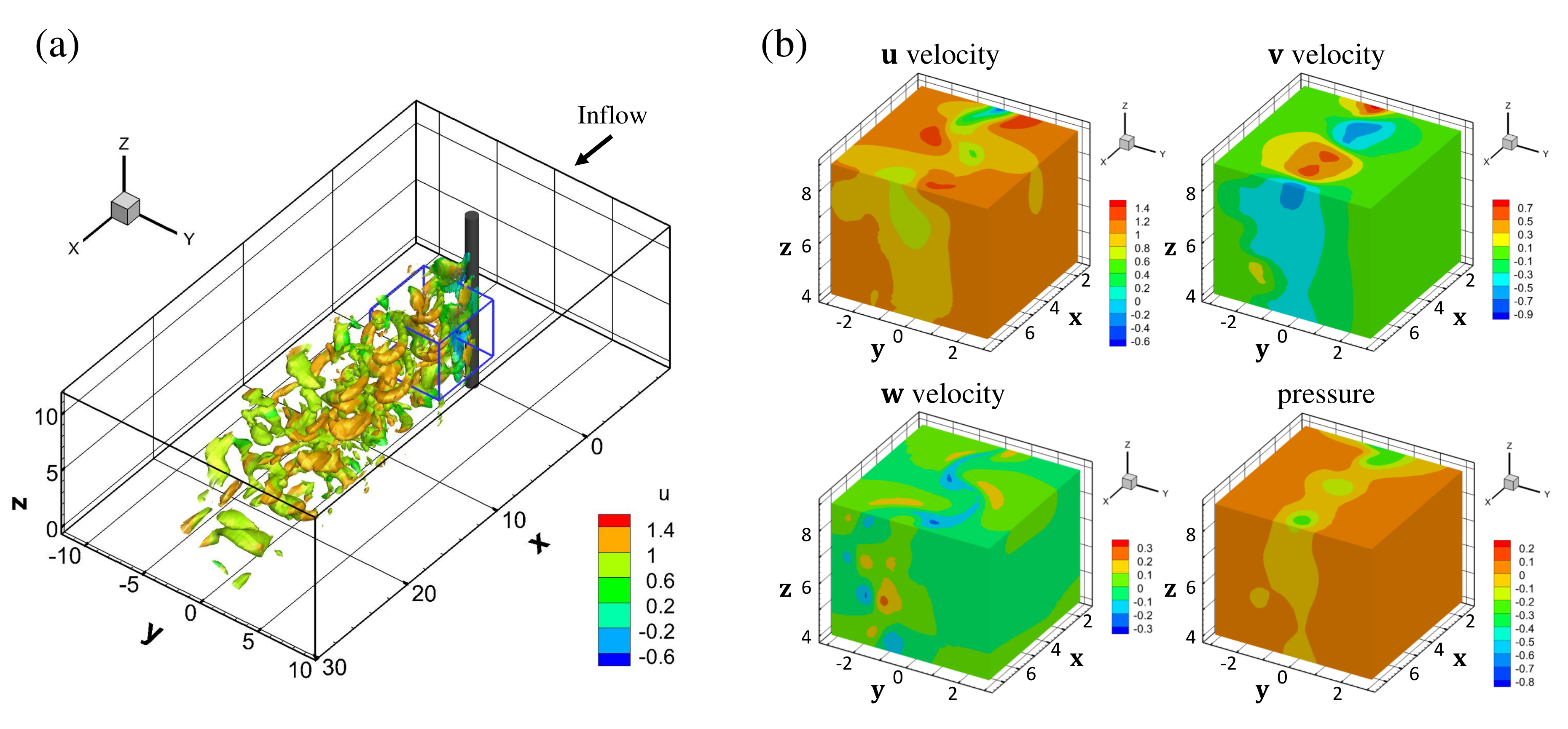}
\caption{\textbf{Case study of PINNs for incompressible flows: illustration of simulating the 3D wake flow over a circular cylinder.} (a) Iso-surface of the vorticity (x-component) in the whole domain color-coded by the streamwise velocity. The cube with blue edges represents the computational domain in this case. (b) Velocity and pressure fields in the domain. The simulation was performed by the CFD solver Nektar, which is based on the spectral/{\em hp} element method~\cite{GK_CFDbook}.
}
\label{fig:incomp_simulation}
\end{figure*}
% ==================================

\subsection{Recent Advances of PINNs}

First proposed in~\cite{raissi2017physicsA,raissi2017physicsB}, see also \cite{raissi2019physics}, PINNs have attracted a lot of attention in the scientific computing community as well as the fluid mechanics community. Here, we review some related works regarding the methodology and the application to fluid mechanics. 

Beneficial due to the high flexibility and the expressive ability in function approximation, PINNs have been extended to solve various classes of PDEs, e.g., integro-differential equations~\cite{pang2019fpinns}, fractional equations~\cite{pang2019fpinns}, surfaces PDEs~\cite{fang2019physics} and stochastic differential equations~\cite{zhang2020learning}. A variational formulation of PINNs based on the Galerkin method (hp-VPINN) was proposed to deal with PDEs with non-smooth solutions~\cite{kharazmi2021hp}. 
In addition, the variational hp-VPINN considered domain decomposition, and similar pointwise versions were also studied in CPINN~\cite{jagtap2020conservative}, and XPINN~\cite{jagtap2020extended}. A general parallel implementation of PINNs with domain decomposition for flow problems is presented in \cite{shukla2021parallel}; the NVIDIA library SimNet~\cite{hennigh2020nvidia} is also a very efficient implementation of PINNs. 
Another important extension is the uncertainty quantification for the PDE solutions inferred by neural networks~\cite{yang2019adversarial,zhang2019quantifying,ZhuZabKouPer2019JCP,sun2020physics,yang2021bpinns}. This has been studied by using the Bayesian framework~\cite{yang2021bpinns}. 
Moreover, some other researches on PINNs focused on the development of the neural network architecture and the training, e.g., using multi-fidelity framework~\cite{meng2020composite}, adaptive activation functions~\cite{jagtap2020adaptive} and dynamic weights of the loss function~\cite{wang2020understanding}, hard constraints~\cite{lu2021physics} and CNN-based network architectures~\cite{gao2021phygeonet}, which can improve the performance of PINNs on different problems. 
On the theoretical side, some recent works~\cite{shin2020convergence,mishra2020estimates,wang2020when} have provided more guarantees and insights into the convergence of PINNs.

The development of the methodology has inspired a number of applications in other fields, especially in fluid mechanics where the flow phenomena can be described by the NSE. In~\cite{raissi2019physics}, the vanilla PINN was proposed to infer the unknown parameters (e.g., the coefficient of the convection term) in the NS equations based on velocity measurements for the 2D flow over a cylinder. Following this work, PINNs were then applied to various flows~\cite{raissi2019deep,jin2021nsfnets,raissi2020hidden,mao2020physics,yin2021non,kissas2020machine,yang2019predictive,lou2020physics,cai2021physics,cai2021flow,wang2021deep,lucor2021physics,mahmoudabadbozchelou2021data}, covering the applications on compressible flows~\cite{mao2020physics}, biomedical flows~\cite{yin2021non,kissas2020machine,arzani2021uncovering}, turbulent convection flows~\cite{lucor2021physics}, free boundary and Stefan problems~\cite{wang2021deep}, etc. The main attractive advantage of PINNs in solving fluid mechanics problems is that a unified framework (shown in Fig.~\ref{fig:schematic}) can be used for both forward and inverse problems. Compared to the traditional CFD solvers, PINNs are superior at integrating the data (observations of the flow quantities) and physics (governing equations). A promising application is on the flow visualization technology~\cite{raissi2020hidden,cai2021artificial}, where the flow fields can be easily inferred from the observations such as concentration fields and images. On the contrary, such inverse problems are difficult for conventional CFD solvers. More relevant works on mechanics in general can be found in \cite{wang2020towards} for turbulent flow, \cite{goswami2020transfer} for phase-field fracture model, and~\cite{zhang2020physics} for inferring modulus in a nonhomogeneous material.

% ================================

\section{Case Study for 3D Incompressible Flows}\label{sec:incomp_flows}

In this section, we demonstrate the effectiveness of PINNs for solving inverse problems in incompressible flows. In particular, we apply PINNs to reconstruct the 3D flow fields based on a few two-dimensional and two-component (2D2C) velocity observations. The proposed algorithm is able to infer the full velocity and pressure fields very accurately with limited data, which is promising for diagnosis of complex flows when only 2D measurements (e.g., planar particle image velocimetry) are available.

% =================================
\begin{figure*}[t]
  \includegraphics[width=0.9\textwidth]{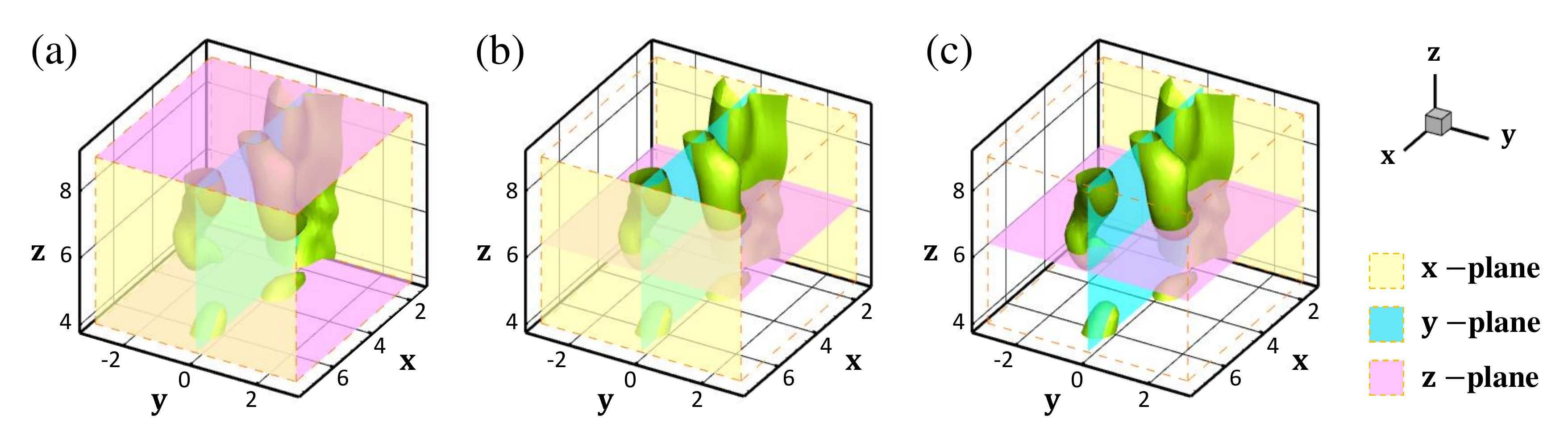}
\caption{\textbf{Case study of PINNs for incompressible flows: problem setup for 3D flow reconstruction from 2D2C observations.} (a) Case 1: two x-planes ($x=1.5, 7.5$), one y-plane ($y=0$) and two z-planes ($z=4.0,9.0$) are observed. (b) Case 2: two x-planes ($x=1.5, 7.5$), one y-plane ($y=0$) and one z-plane ($z=6.4$) are observed. (c) Case 3: one x-plane ($x=1.5$), one y-plane ($y=0$) and one z-plane ($z=6.4$) are observed. Note that for the cross-planes, only the projected vectors are measured. The goal is to infer the 3D flow in the investigated domain using PINNs from these 2D2C observations. 
}
\label{fig:incomp_setup}
\end{figure*}
% =================================

\subsection{Problem setup}

We consider the 3D wake flow past a stationary cylinder at Reynolds number $\text{Re}=200$ in this section. In order to evaluate the performance of PINNs, we generate the reference solution numerically by using the spectral/{\em hp} element method \cite{GK_CFDbook}. 
The computational domain is defined as $\Omega$: $[-7.5,28.5]\times[-20,20]\times[0,12.5]$, where the coordinates are non-dimensionalized by the diameter of the cylinder. The center of the cylinder is located at $(x,y)=(0,0)$. %
We assume that the velocity ($u = 1$) is uniform at the inflow boundary where $x=-7.5$. A periodic boundary condition is used at the lateral boundaries where $y=\pm 20$, and the zero-pressure is prescribed 
at the outlet where $x = 28.5$. Moreover, the no-slip boundary condition is imposed on the cylinder surface. The governing equations in this case study are the dimensionless incompressible Navier-Stokes equations with $\text{Re}=200$. Under this configuration, the wake flow over the cylinder is 3D and unsteady. 
The simulation is performed until the vortex shedding flow becomes stable. At the end, the time-dependent data are collected for PINN training and evaluation. 

The simulation results of the 3D flow are shown in Fig.~\ref{fig:incomp_simulation}, where Fig.~\ref{fig:incomp_simulation}(a) shows the iso-surface of streamwise vorticity ($\omega_x=-0.3$) color-coded with the streamwise velocity $u$. In this section, we are interested in the 3D flow reconstruction problem from limited data, and we only focus on a sub-domain in the wake flow, namely $\Omega_s$: $[1.5,7.5]\times[-3,3]\times[4,9]$, which is represented by a cube with blue edges in Fig.~\ref{fig:incomp_simulation}(a). The contours of the three velocity components and pressure field are shown in Fig.~\ref{fig:incomp_simulation}(b). 
An Eulerian mesh with $61\times61\times26$ grid points is used for plotting. To demonstrate the unsteadiness of the motion, we consider 50 snapshots with $\Delta t=0.2$, which is about two periods of the vortex shedding cycle.

Here, we aim to apply PINNs for reconstructing the 3D flow field from the velocity observations of a few 2D planes. As illustrated in Fig.~\ref{fig:incomp_setup}, three different ``experimental" setups are considered in this paper:
\begin{itemize}
\item Case 1: two x-planes ($x=1.5, 7.5$), one y-plane ($y=0$) and two z-planes ($z=4.0,9.0$) are observed.
\item Case 2: two x-planes ($x=1.5, 7.5$), one y-plane ($y=0$) and one z-plane ($z=6.4$) are observed. 
\item Case 3: one x-plane ($x=1.5$), one y-plane ($y=0$) and one z-plane ($z=6.4$) are observed.
\end{itemize}
We note that for these cross-planes, only the projected vectors (two components) are considered known. For example, the velocities $(u,v)$ can be observed on the z-plane, while the orthogonal component $(w)$ is unknown. The purpose of doing this is to mimic the planar particle image velocimetry in real experiments. Moreover, the resolutions of these 2D2C observations are different, which 
can be found in Table~\ref{tab:incomp_data}.

%%%%%%%%%%%%%%%%%%%%%%%%%%%%%
\begin{table}[t]
\centering
\caption{Case study of PINNs for incompressible flows: details of the 2D2C observations.}\label{tab:incomp_data}
\begin{tabular}{|c|c|c|}
  \hline
  \multirow{2}{*}{Cross-section} &  Observed  &  Observed  \\
  & velocity components  &  spatial resolution  \\
  \hline
  x-plane  &  $(v,w)$   &  $61\times26$  \\
  y-plane  &  $(u,w)$  &  $61\times26$ \\
  z-plane  &  $(u,v)$  &  $61\times61$  \\
  \hline
\end{tabular}
\end{table}
%%%%%%%%%%%%%%%%%%%%%%%%%%%%%

% =================================
% Figure 4
% =================================
\begin{figure*}[t]
  \includegraphics[width=\textwidth]{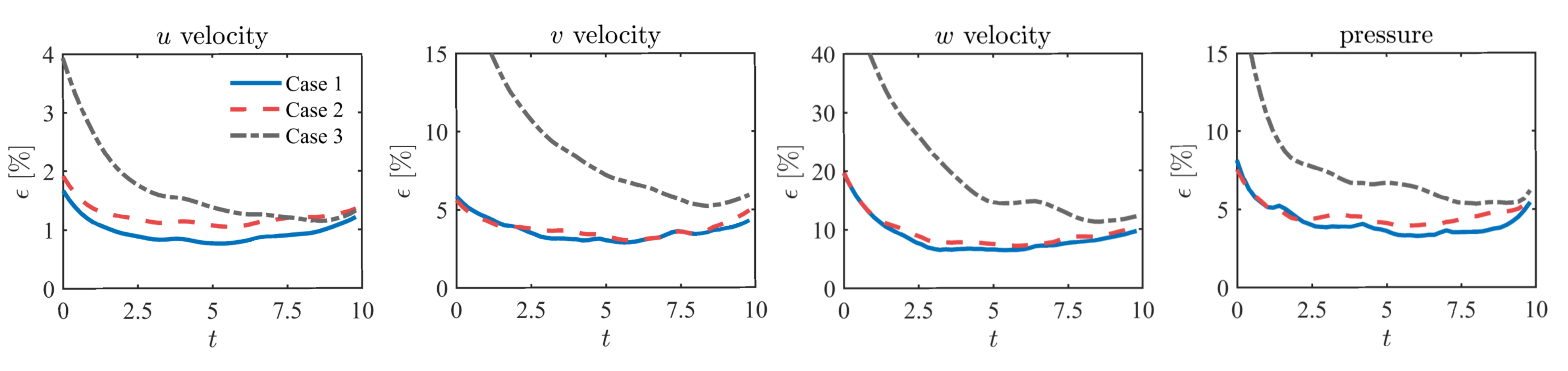}
\caption{\textbf{Case study of PINNs for incompressible flows: relative $L_2$-norm errors of velocities and pressure for different flow reconstruction setups.} These three cases correspond to those shown in Fig.~\ref{fig:incomp_setup}. The errors are computed over the entire investigated domain.
}
\label{fig:incomp_errors}
\end{figure*}
% =================================

\subsection{Implementation of PINNs}

Given the observation data on the cross-planes, we train a PINNs model to approximate the flow fields over the space-time domain. %
The PINN in this section take $(\mathbf{x},t) = (x,y,z,t)$ as inputs and outputs the velocity and pressure $(u,v,w,p)$. 
The loss function in PINN can be defined as:
\begin{equation}
{L} = {L}_{data} + {L}_{PDE},
\end{equation}\label{eq:incomp_loss}
where
\begin{equation}
\begin{aligned}
{L}_{data} =& \frac{1}{N_u} \sum_{i}^{N_u} \parallel u(\mathbf{x}^{i}_{data},t^{i}_{data})-u^{i}_{data} \parallel^{2}  \\
& + \frac{1}{N_v} \sum_{i}^{N_v} \parallel v(\mathbf{x}^{i}_{data},t^{i}_{data})-v^{i}_{data} \parallel^{2} \\
& + \frac{1}{N_w} \sum_{i}^{N_w} \parallel w(\mathbf{x}^{i}_{data},t^{i}_{data})-w^{i}_{data} \parallel^{2} ,
\end{aligned}
\end{equation}
and
\begin{equation}
\begin{aligned}
{L}_{PDE} =& \frac{1}{N_f} \sum_{i}^{N_f} \sum_{j}^{4} \parallel f_{j}(\mathbf{x}^{i}_{f},t^{i}_{f}) \parallel^{2},
\end{aligned}\label{eq:incomp_loss_PDE}
\end{equation}
\begin{equation*}
\begin{aligned}
f_{1,2,3} &= \frac{\partial \mathbf{u}}{\partial t} + (\mathbf{u}\cdot \nabla)\mathbf{u} + \nabla p - \frac{1}{\text{Re}}\nabla^{2}\mathbf{u}, \\
f_{4} &= \nabla \cdot \mathbf{u}.
\end{aligned}
\end{equation*}
The data loss ${L}_{data}$ is composed of three components, and the number of training data (namely $N_u$, $N_v$ and $N_w$) depends on the number of observed planes, the data resolution of each plane as well as the number of snapshots.  
On the other hand, the residual points for ${L}_{PDE}$ can be randomly selected, and here we sample $N_{f}=3\times10^{6}$ points over the investigated space and time domain $\Omega_s$. 
Note that in this study, the boundary and initial conditions are not required unlike the classical setting. Moreover, no information about the pressure is given. The weighting coefficients for the loss terms are all equal to 1. 

A fully-connected neural network with 8 hidden layers and 200 neurons per layer is employed. The activation function of each neuron is $\sigma=sin(\cdot)$.
We apply the ADAM optimizer with mini-batch for network training, where a batch size of $N=10^{4}$ is used for both data and residual points. The network is trained for 150 epochs with learning rates $1\times10^{-3}$, $5\times10^{-4}$ and $1\times10^{-4}$ for every 50 epochs. After training, the velocity and pressure fields are evaluated on the Eulerian grid for comparison and visualization.

\subsection{Inference results}

For a quantitative assessment, we first define the relative $L_2$-norm error as the evaluation metric, which is expressed as:
\begin{equation}
\epsilon_{V} = \frac{\parallel V_{CFD}-\hat{V}\parallel^{2} }{\parallel V_{CFD}\parallel^{2}} \times 100\%,
\end{equation}
where $V\in \{u,v,w,p\}$; $V_{CFD}$ and $\hat{V}$ account for the CFD data and the output of PINNs, respectively. 
We compute the errors in the investigated time domain, which are shown in Fig.~\ref{fig:incomp_errors}. 
It can be seen from the plots that PINNs can infer the 3D flow very accurately for Case 1 and Case 2; using 5 planes (Case 1) is slightly better than using 4 planes (Case 2). When only 3 cross-planes are available (Case 3), the errors become much larger. However, the result of Case 3 is still acceptable as we are able to infer the main flow features with high accuracy (the error of the streamwise velocity is mostly less than 2\%). 
We note that the errors of $w$-velocity are larger than the other components since the $w$-velocity magnitude is relatively small. 
Moreover, we observe larger discrepancy for the initial and final time instants, which can be
attributed to the lack of training data for computing derivatives at $t<0$ and $t>10$. 
This generally happens in the cases when the initial condition is not provided in the unsteady case~\cite{raissi2020hidden}.

\begin{figure*}[t]
  \includegraphics[width= \textwidth]{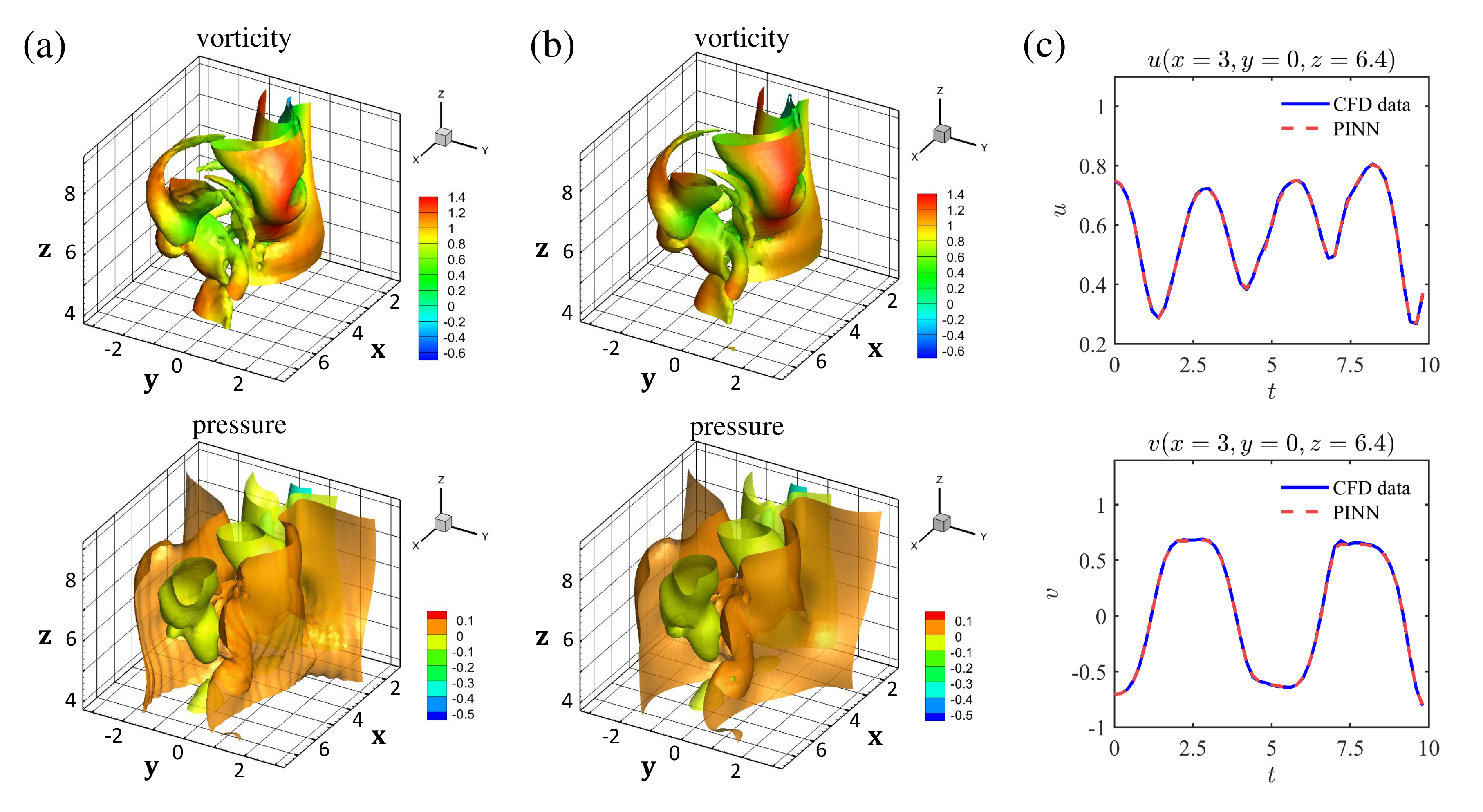}
\caption{\textbf{Case study of PINNs for incompressible flows: inference result of PINNs for Case 2.} (a) Iso-surfaces of vorticity magnitude (top) and pressure (bottom) at $t=8.0$ from CFD data. (b) Iso-surfaces of vorticity magnitude (top) and pressure (bottom) at $t=8.0$ inferred by PINNs. (c) Point measurement $(x=3,y=0,z=6.4)$ of velocity $(u,v)$ against time. In this case, the 3D flow is inferred by PINNs from four cross-planes.%, namely two x-planes ($x=1.5, 7.5$), one y-plane ($y=0$) and one z-plane ($z=6.4$). 
}
\label{fig:incomp_fields}
\end{figure*}

To visualize more details, we emphasize the result of Case~2 and demonstrate the iso-surface of vorticity magnitude and iso-surfaces of pressure at $t=8.0$ in Fig.~\ref{fig:incomp_fields}, where Fig.~\ref{fig:incomp_fields}(a) shows the reference CFD data and Fig.~\ref{fig:incomp_fields}(b) shows the result inferred by PINNs. The vorticity value is $|\omega|=1.2$ and the color represents the streamwise velocity component. It can be seen that the PINNs inference result (inferred from a few 2D2C observations) is very consistent with the CFD simulation. In addition, the velocities $(u,v)$ at a single point $(x=3,y=0,z=6.4)$ against time are plotted in Fig.~\ref{fig:incomp_fields}(c), where we can find that PINNs can capture the unsteadiness of vortex shedding flow very accurately.

% ================================
\section{Case Study for Compressible Flows}
PINNs have also been used in simulating high-speed flows~\cite{mao2020physics}. 
In this section, we consider the following 2D steady compressible Euler equations:
\begin{equation}\label{NCL}
    \nabla\cdot {f}(U) = 0, \; x\in \Omega\subset \mathbb{R}^2,
\end{equation}
where
$$U = [\rho, \rho u, \rho v, \rho E]^T,\; {f} = (G_1, G_2)$$
with 
$G_1(U) =[\rho u,
p+ \rho u^2,
\rho u v,
p u + \rho u E],\;
G_2(U) = [
 \rho v,
\rho uv ,
p+ \rho v^2,
p v+ \rho v E].$
%$$\rho u, $$
% $$G_1(U) =[
% \rho u,
% p+ \rho u^2,
%  \rho u v,
%  p u+ \rho u E],
%  $$
%  \; 
%  G_2(U) = [
%  \rho v,
% \rho uv ,
% p+ \rho v^2,
% p v+ \rho v E].
% $$ U =  \left( {\begin{array}{c}
% \rho\\
% \rho u\\
% \rho v\\
% %\rho u_3\\
% \rho E\\
% \end{array} } \right),
% \; {f} = (G_1, G_2), \;
% \text{with }
% G_1(U) =\left( {\begin{array}{c}
% \rho u\\
% p+ \rho u^2 \\
%  \rho u v\\
% %\delta_{i3}p+ \rho u_3u_i\\
%  p u+ \rho u E\\
% \end{array} } \right), 
% \;
% G_2(U) =\left( {\begin{array}{c}
% \rho v\\
% \rho uv \\
% p+ \rho v^2\\
%  p v+ \rho v E\\
% \end{array} } \right).
% $$
Here, $\rho$ is the density, $p$ is the pressure, $[u,\,v] $ are the velocity components,  and $E$ is the total energy.
% %
We use the additional equation of state, which describes the relation of the pressure and energy, to close the above Euler equations. For instance, we consider the equation of state for a polytropic gas given by
\begin{equation}\label{eq:EOS}
    p = (\gamma-1)\left(\rho E - \frac{1}{2}\rho \|{\bm u}\|^2\right),
\end{equation}
where $\gamma$ is the adiabatic index and $\bm{u} = (u, v)$.

We shall employ the PINNs to solve the inverse problem of the compressible Euler equ.~\ref{NCL}. In particular, we shall infer the density, pressure and velocity fields by using PINNs based on the information of density gradients, limited data of pressure (pressure on the surface of the body), inflow conditions and global physical constrains. 

% ================================
% Figure 6
% ================================
\begin{figure*}[t]
  \centering
    \subfloat{
    \includegraphics[width=0.4\textwidth]{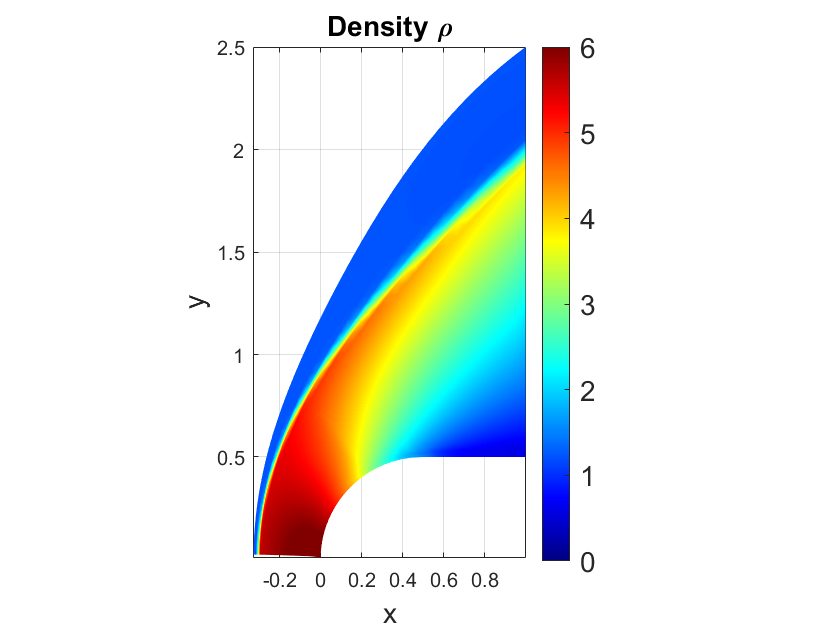}
    }
    \subfloat{
    \includegraphics[width=0.4\textwidth]{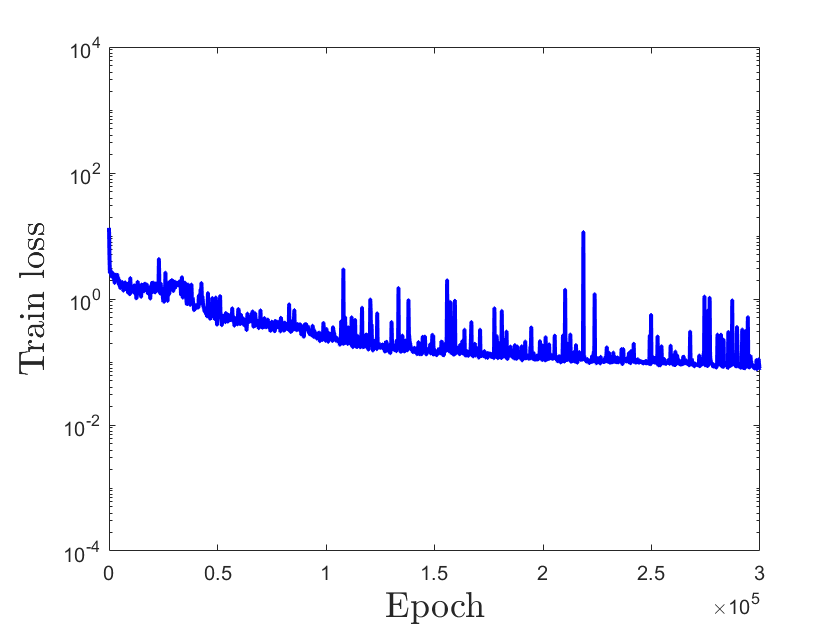}
    }
\caption{\textbf{Case study of PINNs for compressible flows.} Left: the density obtained by using CFD simulation with the inlet flow condition \eqref{inflow}. Right: training loss vs. number of epochs.
}
\label{fig:comp_setup:loss}
\end{figure*}
% ================================
% ================================
% Figure 7
% ================================
\begin{figure}[t]
  \centering
    % \subfigure{
    % \includegraphics[width=0.26\textwidth]{Figs_comp/loss_Bowshock_dw.png}
    % }
    % \subfloat{
    % \includegraphics[width=0.42\textwidth]{Figs_comp/flat_p_uvout.png}
    % } \\
    % \subfloat{
    % \includegraphics[width=0.42\textwidth]{Figs_comp/flat_u_uvout.png}
    % }
    \includegraphics[width=0.48\textwidth]{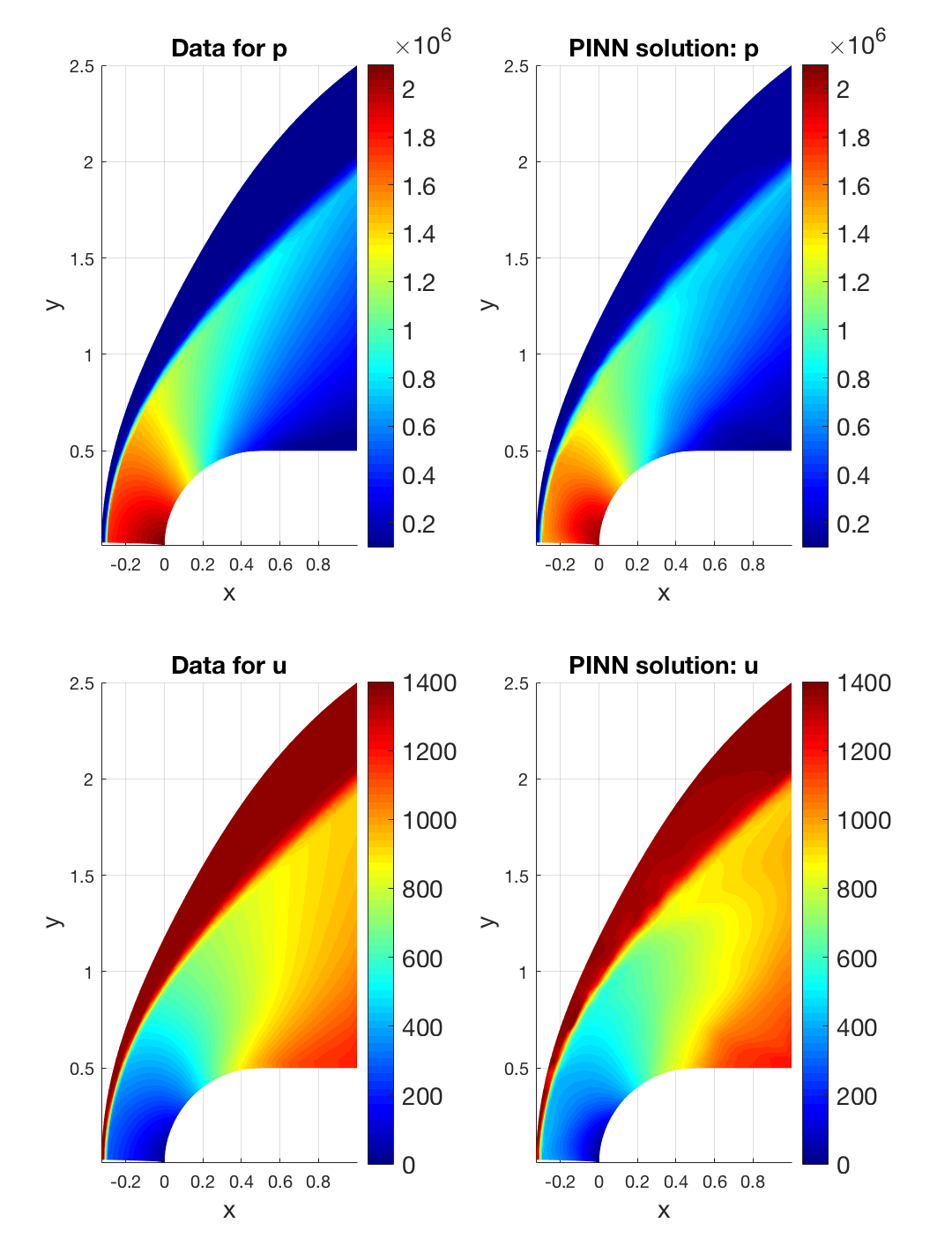}
\caption{\textbf{Case study of PINNs for compressible flows.} Comparison between the PINN solutions and the CFD solutions. Top: pressure $p$, Bottom: velocity component $u$.
}
\label{fig:comp:p:u}
\end{figure}
% ================================

\subsection{Problem setup}
We consider a 2D bow shock wave problem.
For traditional CFD simulations, it is crucial that the boundary conditions, which play an important role, are properly implemented. However, for the shock wave problem in high-speed flows, the boundary conditions in real experiments  are usually not known and can only be estimated approximately. 
In the present work, instead of using most of the boundary conditions required by the traditional CFD simulation, we solve the Euler equ.~\ref{NCL} using PINNs based on the data of density gradients $\nabla \rho$ motivated by the Schlieren photography available experimentally; additionally, we use limited data of the surface pressure obtained by pressure sensors as well as the global constrains (mass, momentum and energy). 
We also use the inflow conditions here. Note that unlike the one-dimensional case, where the density gradient in the whole computational domain is used in~\cite{mao2020physics}, here we only use the density gradients in a sub-domain $D$ of the computational domain $\Omega$, i.e., $D \subset \Omega$.
By combining the mathematical model and the given data, we have the weighted loss function of PINN given by
\begin{equation}\label{loss:drho:p:inlet:U}
\begin{aligned}
    Loss  =  \omega_{1}Loss_{F} + \omega_{2}Loss_{\nabla \rho|_{D}} + \omega_{3}Loss_{inflow}  + \omega_{4}Loss_{p^*} &\\
    + \omega_{5}\left(Loss_{Mass} + Loss_{Momentum} + Loss_{Energy}\right) 
    + \omega_{6}Loss_{n\cdot {\bm u}},&
\end{aligned}
\end{equation}
whre the last term corresponds to the velocity condition on the surface.
% where $Loss_{n\cdot {\bm u}} = \frac{1}{N_{\bm u}} \sum_{k = 1}^{N_{\bm u}} (n\cdot {\bm u})^2(x_k, y_k)$ and $(x_k, y_k),\; k= 1,\ldots, N_{\bm u}$ are points located on the surface of the body.

\subsection{Inference results}
To demonstrate the effectiveness of PINNs for compressible flows,
% we consider the following example.
% \begin{exam}\label{ex:bow}
we consider the bow shock problem with the following inlet flow conditions 
\begin{equation}\label{inflow}
\begin{aligned}
    &M_\infty = 4,\; p_\infty = 101253.6Pa,\; \rho_\infty = 1.225kg/m^3,\\
    &u_\infty = 1360.6963m/s,\; v_\infty = 0,\; T_\infty = 288K.
\end{aligned}
\end{equation}
  %  $M_\infty = 4,\; p_\infty = 101253.6Pa,\; \rho_\infty = 1.225kg/m^3,\; u_\infty = 1360.6963m/s,\; v_\infty = 0,\; T_\infty = 288K.$
%The distributions of the residual points, the data points for the density gradient and the inflow conditions are shown in Fig. \ref{fig:distribution:bow:h}. 
The data points for the pressure are located on the surface of the body.
% \end{exam}
%
By using the above inflow conditions and CFD code, we can obtain the steady state flow. We show the density computed by CFD in the left plot of Fig. \ref{fig:comp_setup:loss}. We employ a $6\times 60$ (6 hidden layer) neural network and train it by using layer-wise adaptive tanh activate function~\cite{jagtap2020adaptive} and the Adam optimizer with the learning late being $6\times 10^{-4}$ and $3\times 10^{5}$ epochs. Here, we also use the technique of dynamic weights~\cite{wang2020understanding,jin2021nsfnets}.
The history of the training loss is shown in the right plot of Fig. \ref{fig:comp_setup:loss}. The results of the PINN solutions for the pressure and velocity ($u$) are shown in Fig.~\ref{fig:comp:p:u}. Observe that the PINN solutions are in good agreement with the CFD data. This indicates that we can reconstruct the flow fields for high-speed flows using some other available knowledge different from the boundary conditions required by the traditional CFD simulation.

% =============================
% Figure 8
% =============================
\begin{figure*}[t]
    \centering
	\includegraphics[width=0.9\textwidth]{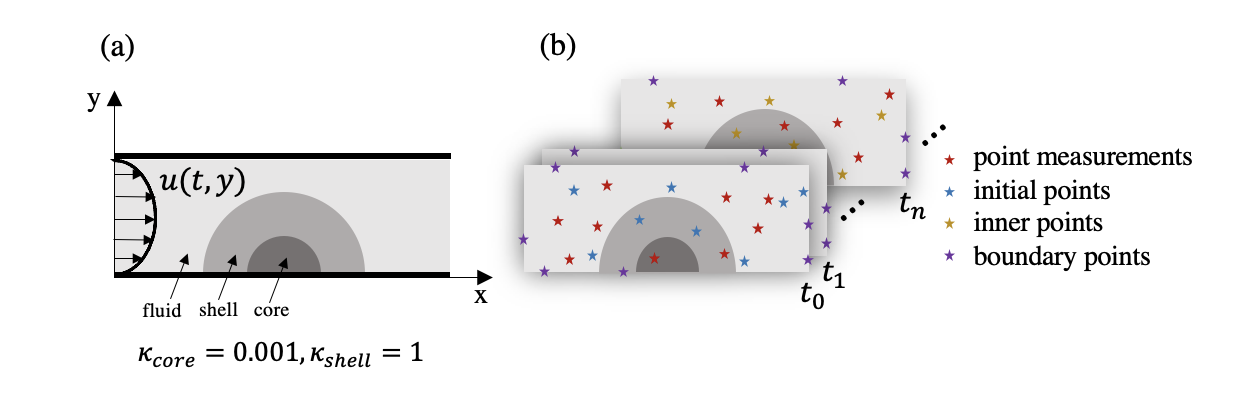}
    \caption{\textbf{Case study for 2D flow past a thrombus with phase dependent permeability.} (a) A channel with walls on the top and bottom boundaries with the inlet flow $u(t,y)$ entering from the left side; $\phi$=1 is the fluid. A thrombus with an impermeable core $\phi=-1$ and permeable shell $\phi=0$ is present at the bottom boundary. (b) Sampling points for inferring permeability include initial points (\textcolor{ProcessBlue}{$\star$}) at the time $t_{0}$, inner points (\textcolor{Dandelion}{$\star$}) from $t_{1}$ to $t_{n}$, boundary points (\textcolor{Plum}{$\star$}) on boundaries, and point measurements (\textcolor{BrickRed}{$\star$}) with PDE solutions. (Figure adapted from \cite{yin2021non}).}
    \label{fig:throm_setup}
\end{figure*}
% =============================

% ================================
\section{Case Study for Biomedical Flows}
% \subsection{Model}
In addition to the aforementioned flow examples, PINNs have also been used in biomedical flows~\cite{yin2021non}. In this section, we consider inferring material properties of a thrombus in arterial flow described by the Navier-Stokes and Cahn-Hilliard equations. Such equations can be used for describing the mechanical interaction between thrombus and blood flow as a fluid-structure interaction (FSI) problem. The PDE system can be written as:
\begin{align}
    \rho(\frac{\partial \mathbf{u}}{\partial t} + \mathbf{u}\cdot\nabla\mathbf{u}) +\nabla p &= \nabla \cdot (\bm{\sigma_{vis}} + \bm{\sigma_{coh}}) - \mu\frac{(1-\phi)\mathbf{u}}{2\kappa(\phi)}, \label{ns_ch}\\
    \nabla\cdot \mathbf{u} &= 0, \\
    \frac{\partial \bm{\psi}}{\partial t} + \mathbf{u}\cdot \nabla\bm{\psi} &= 0, \\
    \frac{\partial \phi}{\partial t} + \mathbf{u}\cdot \nabla \phi &= \tau \Delta \omega,\\
    \omega &= \Delta \phi + \gamma g(\phi),
\end{align}
$g(\phi)$ is the derivative of the double-well potential $(\phi^{2}-1)^{2}/4h^{2}$. $\mathbf{u}(\mathbf{x}, t)$, $p(\mathbf{x}, t)$ $\bm{\sigma}(\mathbf{x}, t)$, and $\phi(\mathbf{x}, t)$ represent the velocity, pressure, stress, and phase field. $h$ is the interfacial length; $\psi$ denotes the auxiliary vector and its gradients are the components of the deformation gradient tensor $\mathbf{F}$ as follows:
\begin{equation}
\nonumber
    \mathbf{F} := \begin{bmatrix}
            -\frac{\partial \psi_{1}}{\partial y} & -\frac{\psi_{2}}{\partial y}\\
            \frac{\partial \psi_{1}}{\partial x} & \frac{\partial \psi_{2}}{\partial x}
            \end{bmatrix}.
\end{equation}
Equ.~\ref{ns_ch} is the Navier-Stokes equation with viscous and cohesive stresses, respectively, which can be written as:
\begin{align}
    &\bm{\sigma_{vis}} = \mu \nabla u, \\
    %&\bm{\sigma_{ela}} = \nabla\cdot(\lambda_{e}\frac{(1-\phi)}{2}(\mathbf{F}\mathbf{F}^{T}-\mathbf{I})),\\
    &\bm{\sigma_{coh}} = \lambda\nabla\cdot(\nabla\phi \otimes \phi).
\end{align}
Moreover, $\gamma$, $\tau$, and $\lambda$ are the interfacial mobility, relaxation parameter, and mixing energy density, respectively. We follow the normalization in~\cite{wufsus2013hydraulic} and write $\kappa = \frac{k_{f}}{a_{f}^{2}}$, $k_{f}$ is the true permeability and $a_{f}$ is the fibrin radius. We set the density $\rho$ = 1, viscosity $\mu$ = 0.1, $\lambda$ = $4.2428\times10^{-5}$, $\tau$ = $10^{-6}$, and the interface length $h$= 0.05. These parameters in PINNs are non-dimensionalized values so as to be consistent with the CFD solver. 

The velocity at inlet $\Gamma_{i}$ is set as Dirichlet boundary conditions $\textbf{u} = g, (\textbf{x}, t) \in \Gamma_{i} \times (0, T)$. We impose the no-slip boundary on the wall $\Gamma_{w}$ and Neumann boundary conditions, i.e., $\frac{\partial \phi}{\partial \textbf{n}} = \frac{\partial \omega}{\partial \textbf{n}} = 0, \textbf{x} \in \Gamma_{w} \cup \Gamma_{i} \cup \Gamma_{o}$ for $\phi$ and $\omega$ at all boundaries.

\subsection{PINNs}
We construct two fully-connected neural networks, \textit{Net U} and \textit{Net W}, where the outputs of \textit{Net U} represents a surrogate model for the PDE solutions \textit{u, v, p}, and \textit{$\phi$} and the outputs of \textit{Net W} are PDE solutions \textit{$\omega$, $\psi_{1}$}, and \textit{$\psi_{2}$}. Each network has 9 hidden layers with 20 neurons per layer. The total loss \textit{L} is a combination of different losses as:
\begin{equation}
    L = \omega_{1}L_{PDE} + \omega_{2}L_{IC} + \omega_{3}L_{BC} + \omega_{4}L_{data},
\end{equation}
where \textit{$L_{PDE}$} is the PDEs residual loss, \textit{$L_{IC}$} is the initial condition loss,  \textit{$L_{BC}$} is the boundary condition loss, and \textit{$L_{data}$} is the data loss. In particular:
\begin{align}
    L_{PDE}(\theta, \bm{\lambda}; X_{PDE}) &= \frac{1}{ \left| X_{PDE} \right| } \sum\limits_{\textbf{x}\in X_{PDE}} \norm{f(\textbf{x}, \partial_{t}{\hat{\textbf{u}}}, \partial_{x}\hat{\textbf{u}},...;\bm{\lambda})}_{2}^{2},\\
    L_{BC}(\theta, \bm{\lambda}; X_{BC}) &= \frac{1}{\left| X_{BC} \right|} \sum\limits_{\textbf{x}\in X_{BC}}\norm{\mathfrak{B(\hat{\textbf{u}}, \textbf{x})}}_{2}^{2},\\
    L_{IC}(\theta, \bm{\lambda}; X_{IC}) &= \frac{1}{\left| X_{IC} \right|} \sum\limits_{\textbf{x}\in X_{IC}}\norm{\hat{\textbf{u}} - \textbf{u}_{t_{0}}}_{2}^{2}, \\
    L_{data}(\theta, \bm{\lambda}; X_{data}) &= \frac{1}{| X_{data} |} \sum\limits_{\textbf{x}\in X_{data}}\norm{\hat{\textbf{u}} - \textbf{u}_{data}}_{2}^{2},
\end{align}
where $\omega_{1}$, $\omega_{2}$, $\omega_{3}$, and $\omega_{4}$ are the weights of each term. The training sets $X_{PDE}$, $X_{BC}$, and $X_{IC}$ are sampled from the inner spatio-temporal domain, boundaries, and initial snapshot, respectively. $X_{data}$ is the set that contains sensor coordinates and point measurements; $\left| \cdot \right|$ denotes the number of training data in the training set. In particular, $\mathfrak{B}$ represents a combination of the Dirichlet and Neumann residuals at boundaries. Finally, we optimize the model parameters $\bm{\theta}$ and the PDE parameters $\bm{\lambda}=[\kappa]$ by minimizing the total loss $L(\bm{\theta}, \bm{\lambda})$ iteratively until the loss satisfies the stopping criteria. Minimizing the total loss is an optimization process for $\bm{\lambda}$ such that the outputs of the PINN satisfy the PDE system, initial/boundary conditions, and point measurements. 
%We use the mean relative $L_{2}$ error ($\epsilon$), same as in~\cite{raissi2020hidden}, to quantify errors between reference data and model predictions:
%\begin{equation}
%    \epsilon := (\frac{1}{N}\sum_{i}^{N}[\hat{u}(\textbf{x}_{i})-u(\textbf{x}_{i})]^{2})/(\frac{1}{N}\sum_{i}^{N}[u(\textbf{x}_{i})-\frac{1}{N}\sum_{i}^{N}u(\textbf{x}_{i})]^{2})
%\end{equation}
%%%%%%%%%%%%%%%%%%%%%%%%%%%%%%%

\subsection{Problem setup}

\begin{figure}[t]
    \centering
	\includegraphics[width=0.5\textwidth]{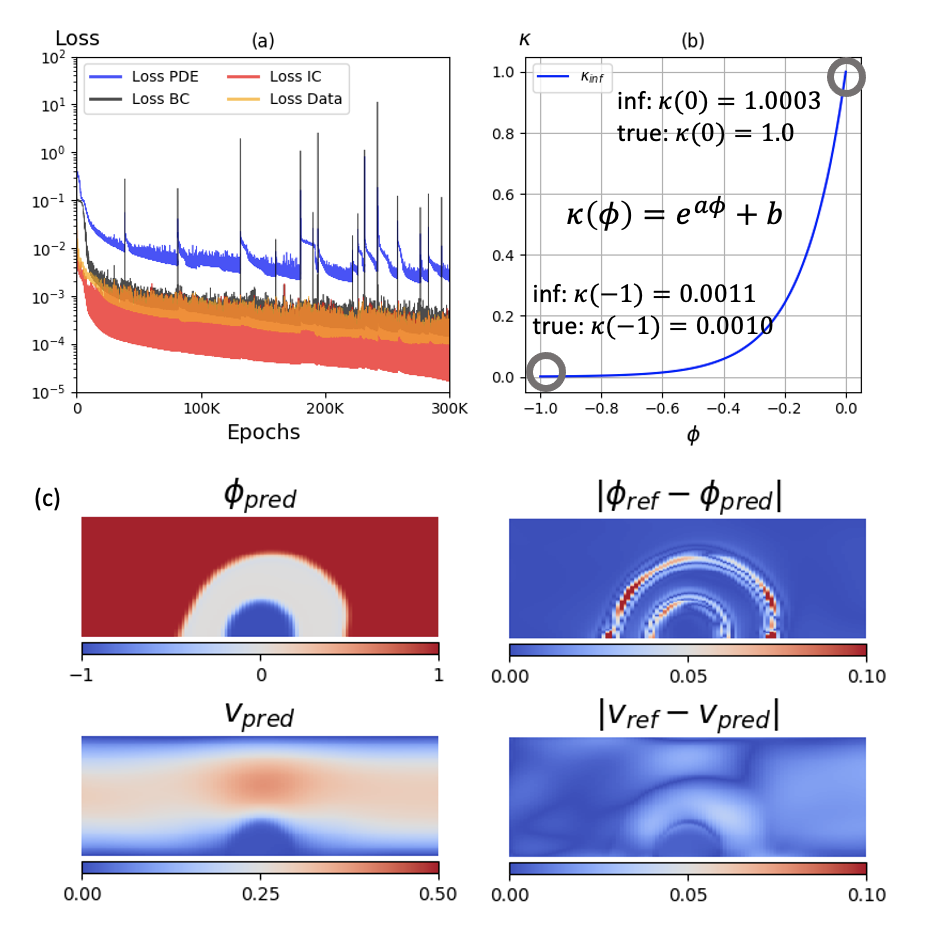}
    \caption{\textbf{Inference results for 2D flow past a thrombus with phase-dependent permeability.} (a) History of network losses (Loss PDE, Loss IC, Loss BC, and Loss Data) and (b) inferred permeability $\kappa$ as a function of $\phi$. (c) Comparison of phase field and velocity field for $\kappa(\phi)$ at $t$ = 0.78 and their absolute error. The core permeability is 0.001 and the shell permeability is set 1 as the actual values. 10,000 data points are scattered in the spatio-temporal domain from 30 snapshots ($t \in [0.03, 0.93]$) as the training data to infer the permeability. Inferred $a$ and $b$ are 7.10 and 0.0003 compared to the true value at 6.9 and 0.0. (Figure adapted from \cite{yin2021non})}
    \label{fig:throm_comp}
\end{figure}

We consider a computational setup with a semi-circle permeable thrombus in a channel with a steady parabolic flow coming from the left (Fig. \ref{fig:throm_setup}). This setup is meaningful in a sense that it presents an idealized thrombus with an impermeable core, which consists of a fibrin clot and a permeable shell (consisting of loosely-packed and partially-activated platelets). This model has been adapted as an idealized thrombus in previous works \cite{zheng2020three,xu2008multiscale,yazdani2017general}. The goal is to infer the unknown permeability (for the shell and core) and velocity field based on the measurable phase field data. Fig. \ref{fig:throm_setup}(b) presents all the types of training data, namely the initial snapshot $t_{0}$ (\textcolor{ProcessBlue}{$\star$}), inner spatio-temporal domain from $t_{1}$ to $t_{n}$ (\textcolor{Dandelion}{$\star$}), and at the boundaries (\textcolor{Plum}{$\star$}). Also, point measurements (\textcolor{BrickRed}{$\star$}) including their coordinates and phase field value are sampled in the spatio-temporal domain to calculate the data loss term in the total loss. We draw 1,000 points from an initial snapshot, 10,000 inner points to compute the PDE residuals, and 1,000 boundary points to compute the residuals at boundaries. Note that point measurements and inner points are drawn from the inner spatio-temporal domain; the former contains the PDE solutions whereas the latter does not.

The thrombus is present in the middle of the channel as shown in Fig. \ref{fig:throm_setup}(a) with permeability $\kappa=0.001$ in the core ($\phi=-1$) and $\kappa=1$ in the outer shell layer ($\phi=0$). To express such spatial variation explicitly, we consider a relation between $\phi$ and $\kappa$ in this case:
\begin{equation}
    \kappa(\phi) = e^{a\phi} + b,
\end{equation}
where $a$ and $b$ are model parameters to be optimized in the PINN model and the true values of $a$ and $b$ are 6.90 and 0.0. Notice that the relation is not unique in terms of its form as long as the permeability value matches the true value for $\phi=0$ and $1$.

\subsection{Inference Results}
We present the inference results in Fig. \ref{fig:throm_comp}. Fig. \ref{fig:throm_comp}(a) shows the history of the different losses, namely PDE loss, boundary condition loss (Loss BC), initial condition loss (Loss IC), and data loss (Loss Data) in (a). We trained the model with 300,000 epochs with PDE loss as the largest component. The other errors are lower than $O(10^{-3})$ Plot (b) shows the inference result for $\kappa$ as a function of $\phi$; the inferred $a$ and $b$ are at 7.1 and 0.0003, indicating that the permeability at core area ($\kappa(\phi=-1)=0.0011$) and shell area ($\kappa(\phi=0)=1.0003$) match the reference values well. As a qualitative comparison, we present the predicted fields of $\phi_{pred}$ and $v_{pred}$ and their difference with the reference data at $t=0.78$ in Fig. \ref{fig:throm_comp}(c). The phase field prediction exhibits a good agreement compared with the ground truth data. More importantly, the hidden velocity field can also be accurately inferred only based on the phase field data. Notice that the errors for the phase field are mainly distributed in and around the outlet layer of the thrombus and the errors in velocity field are mainly confined within the shell layer. 

% ================================
\section{Summary}
PINNs offer a new approach to simulating realistic fluid flows, where some data are available from multimodality measurements
whereas the boundary conditions or initial conditions may be unknown. While this is perhaps the  prevailing scenario in practice, existing CFD solvers cannot handle such ill-posed problems and hence one can think of PINNs for fluid problems as a complementary approach to the plethora of existing numerical methods for CFD for idealized problems. 

There are several opportunities for further research, e.g., using PINNs for active flow control to replace expensive experiments and time-consuming large-scale simulations as in~\cite{fan2020reinforcement}, or predict fast the flow at a new high Reynolds number using transfer learning techniques assuming we have available solutions at lower Reynolds numbers. Moreover, a new area for exploration could be the development of closure models for unresolved flow dynamics at very high Reynolds number using the automatic data assimilation method provided by PINNs. Computing flow problems at scale requires efficient multi-GPU implementations in the spirit of data parallel \cite{hennigh2020nvidia} or a hybrid data parallel and model parallel paradigms as in \cite{shukla2021parallel}. The parallel speed up obtained for flow simulations so far is very good, suggesting that PINNs can be used in the near future for industrial complexity problems at scale that CFD methods cannot tackle.

\acknowledgement{The last author (GEK) would like to acknowledge  support by the Alexander von Humboldt fellowship. }

%%%%%%%%%%%%%%%%%%%%%%%%%%%%%%%%%%%%%
%Reference
\bibliographystyle{spmpsci}
\bibliography{ref.bib}

% \begin{thebibliography}{99}
% \bibitem {JW1}Qiao, L.: Variational constitutive updates for strain gradient isotropic
% plasticity. Massachusetts Institute of Technology, Cambridge
% (2009)
% \bibitem {JW2}Moreau, P., Raulic, M., Ping, K.M.Y., et al.: Measurement of the
% size effect in the yield strength of nickel foils. Philos. Mag. Lett.
% 85, 339--C343 (2005)
% \bibitem {JW3}Idiart, M.I., Fleck, N.A.: Size effects in the torsion of thin metal
% wires. Model. Simul. Mater. Sci. {\bf 18}, 015009 (2010) \href{https://doi.org/10.1007/s10546-018-0404-0}{https://doi.org/10.1007/s10546-018-0404-0}%
% \end{thebibliography}

%\end{sloppypar}
\end{document}